\documentclass{ecai}

\usepackage{latexsym}
\usepackage{amssymb}
\usepackage{graphicx}
\usepackage{color}
\usepackage{times}
\usepackage{soul}
\usepackage{graphicx}
\usepackage{amsmath}
\usepackage{amsthm}
\usepackage{algorithm}
\usepackage{algorithmic}
\usepackage{amssymb}
\usepackage{xspace}

\usepackage{latexsym}

\newtheorem{theorem}{Theorem}
\newtheorem{definition}[theorem]{Definition}

\usepackage{amsfonts}

\renewcommand{\vec}[1]{\boldsymbol{#1}}
\renewcommand{\baselinestretch}{1}
\newcommand{\Design}{$\mathsf{HDQF}$\xspace}

\newcommand{\HDadd}{+}
\newcommand{\HDmult}{\odot}

\usepackage{physics}

\begin{document}
\begin{frontmatter}




\title{Hyperdimensional Quantum Factorization}



\author[A]{\fnms{Prathyush}~\snm{Poduval}\footnote{ppoduval@uci.edu}}
\author[B]{\fnms{Zhuowen}~\snm{Zou}\footnote{zhuowez1@uci.edu}}
\author[C]{\fnms{Alvaro}~\snm{Velasquez}\footnote{alvaro.velasquez@darpa.mil}}
\author[D]{\fnms{Mohsen}~\snm{Imani}\footnote{mohseni@uci.edu}}

\address[A,B,D]{University of California, Irvine}
\address[C]{Defense Advanced Research Projects Agency (DARPA)}

\begin{abstract}
This paper presents a quantum algorithm for efficiently decoding hypervectors, a crucial process in extracting atomic elements from hypervectors—an essential task in Hyperdimensional Computing (HDC) models for interpretable learning and information retrieval. HDC employs high-dimensional vectors and efficient operators to encode and manipulate information, representing complex objects from atomic concepts. When one attempts to decode a hypervector that is the product (binding) of multiple hypervectors, the factorization becomes prohibitively costly with classical optimization-based methods and specialized recurrent networks, an inherent consequence of the binding operation. We propose \Design, an innovative quantum computing approach, to address this challenge. By exploiting parallels between HDC and quantum computing and capitalizing on quantum algorithms' speedup capabilities, \Design encodes potential factors as a quantum superposition using qubit states and bipolar vector representation. This yields a quadratic speedup over classical search methods and effectively mitigates Hypervector Factorization capacity issues.
\end{abstract}

\end{frontmatter}

\begin{figure}[h]
    \centering
    \includegraphics[width=\columnwidth]{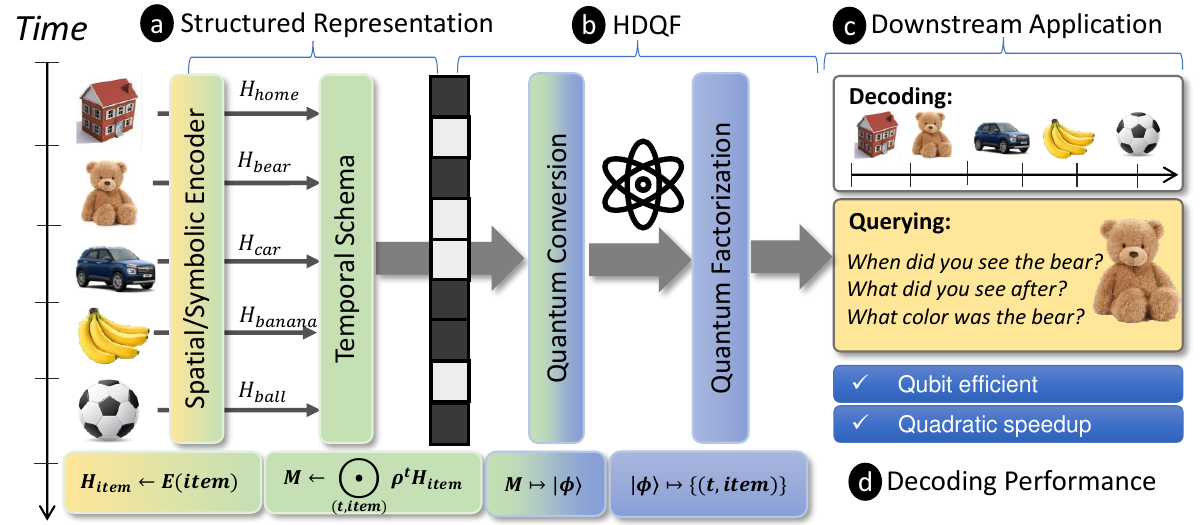}
    \caption{Overview of \Design. (a) HDC Represents a sequence of items across time as a compound hypervector via associative encoding. The items are encoded to item hypervectors, which are then permuted and bound to each other to represent temporal associations. (b) \Design decodes the compound hypervector by first mapping it to quantum states, upon which the modified Grover's algorithm is then applied to recover the correct components. (c) The factorization results can be used for downstream tasks, including retrieving the original information or querying. (d) \Design enables quadratic speedup compared to classical algorithms.}
    \label{fig:Overview}
\end{figure}

\section{Introduction} \label{sec:intro}

Hyperdimensional Computing (HDC), also known as Vector Symbolic Architecture (VSA), employs high-dimensional vectors to encode and process information. In HDC, atomic concepts identified from components of the data are encoded into high-dimensional vectors with brain-inspired properties~\cite{kanerva2009hyperdimensional,ni2023brain,poduval2022adaptive,poduval2021cognitive,imani2022neural}. 
The HDC operators - binding, bundling, and permutation - construct sets, associations, and sequences respectively, facilitating the interpretable creation and manipulation of complex objects for data representation, learning, and processing. 
For learning, an HDC model makes decisions by evaluating the similarity between query and model hypervectors~\cite{kim2018efficient,rahimi2017hyperdimensional2,kleyko2023survey,poduval2021stochd,poduval2021robust}; for cognitive processing, an HDC model retrieves information directly over the hyperspace with HDC operators; it then decodes the information with similarity functions and the atomic hypervectors ~\cite{kanerva2009hyperdimensional,poduval2024nethd,zou2022biohd}. Recent work has shown great advantages of HDC in enhancing the cognitive capability of neural networks in an explainable fashion~\cite{hersche2023neuro}: a neural network learns to encode and perform HDC-like composition of the data over Raven's Progressive Matrix, a visual reasoning task over the symbolic attributes of the objects, and significantly outperforms state-of-the-art pure DNN and neuro-symbolic AI solutions in both accuracy and efficiency scaling. 

While HDC serves as a natural bridge between neural models and symbolic reasoning, it faces some scalability issues when it comes to decoding, a crucial process for information retrieval.
Specifically, when dealing with a hypervector that binds multiple atomic hypervectors, finding its constituents using only the codebooks (lists of atomic hypervectors) becomes computationally demanding~\cite{frady2020resonator,yeung2024self}. 
Consequently, the practical application of VSAs has been limited due to the absence of an efficient solution for accessing elements within compound data structures that associate multiple components.


To address this issue, we opt to harness quantum computing due to the following reasons:
(1) Quantum computing and Hyperdimensional Computing share intriguing similarities, with some HDC approaches emulating functional aspects of quantum computing. Moreover, both approaches leverage the geometry of large state spaces for computation.
(2) Quantum algorithms are renowned for their asymptotic acceleration of specific problems, some even achieving exponential gains. As we found out, our problem falls into this domain.
(3) Classically, factorization algorithm parameters are constrained by problem space and the need for a robust signal-to-noise ratio, driven by HDC's near-orthogonality requirement. In the quantum realm, this limitation diminishes as quantum-generated codebooks can be orthogonal. Recognizing this computational resemblance, certain quantum algorithms may serve as a foundation for tackling the factorization problem.

We present \Design, Hyperdimensional Quantum Factorization, that utilizes quantum computing to perform hypervector factorization. \Design capitalizes on the parallelism between qubit states $0,1$ and bipolar hypervector representation $-1,1$, enabling quantum superposition to embody all potential factorizations. A quantum search algorithm, inspired by Grover's algorithm, then identifies the correct factorization by exploiting quantum systems' capability to concurrently operate on all superposition states. Significantly, \Design yields a quadratic speedup compared to classical search and mitigates capacity concerns. Our contributions are as follows:
\begin{enumerate}
    \item We formally establish the representational connection between HDC and Quantum Computing.
    \item We design \Design, a modified Grover's algorithm specialized for addressing the Hypervector Factorization Problem. 
    \item We implement our quantum algorithm in a quantum circuit model and apply it to several use cases to evaluate its performance. 
\end{enumerate}
In Fig.~\ref{fig:Overview}, we present an overview of the usage of \Design.

\section{Background} \label{sec:HDC}\label{subsec:HDCAtom} \label{subsec:HDCop} \label{subsec:HDCSchema}


In this section, we review Hyperdimensional Computing. We describe each component of HDC by its specifications of desired properties and the implementation most relevant to our task - the Multiply-Add-Permute (MAP) framework \cite{gayler1998multiplicative},

\subsection{Hyperdimensional Computing}

Hyperdimensional Computing uses high-dimensional vectors, called hypervectors, to represent objects. The mapping is typically constrained such that the representation is high dimensional, robust, holistic, and random \cite{kanerva2009hyperdimensional,poduval2021hyperdimensional}, where the similarity $\delta$ between the hypervectors reflects the similarity between the data.
 
In the Multiply-Add-Permute (MAP) VSA~\cite{gayler1998multiplicative}, a datapoint $x$, consisting of discrete set of attributes from a set $\mathcal{D}$, is encoded using a random bipolar hypervector:
\begin{align*}
  h_{x} & \leftarrow \operatorname{Unif}{\{-1, 1\}^{D}} & x \in \mathcal{D} \\
    \delta(h_x, h_y) & = h_x^{T}h_y/D & x, y \in \mathcal{D},
\end{align*}
\noindent where $D$ is the dimension of the hypervector, $\operatorname{Unif}{\{-1, 1\}^{D}}$ is the uniform distribution over the bipolar vectors of dimension $D$, and $\delta(.,.)$ is the similarity metric, which approximates the Kronecker kernel as $D\to\infty$ because $\delta(h_x, h_x) = 1$ and $\delta(h_x, h_y) \sim N(0,1/\sqrt{D})$. We refer to the list of atomic hypervectors $\{h_x\}_{x\in \mathcal{D}}$ as \textit{the codebook}. 

The core operators used for representing data structures in VSA are bundling $\HDadd$, binding $\HDmult$, and permutation $\rho$: 
\begin{enumerate}
\item Bundling creates sets~\cite{kanerva2009hyperdimensional} by ``superposing" the hypervectors such that the resulting hypervector retains similarity with its constituents. In MAP, bundling is the element-wise addition.
\item Binding functions as an association operator such that the resultant contains the information of both of its constituents but is dissimilar to any of them~\cite{frady2018theory}. It is implemented as element-wise multiplication.
\item Permutation enables sequencing of the hypervector by repeated application of a predetermined permutation and is implemented typically as the circular shift. 
\end{enumerate}

The functional aspects of these operators can be easily verified. 
To give an example, if we generate 2 codebooks of size $3$ for ``color" and ``shape" referring to \textit{red, blue, and yellow} and \textit{triangle, square, and circle} respectively. One may represent ``blue circle and yellow square" by $h_{bcys} = h_{blue}\HDmult h_{circle} \HDadd h_{yellow}\HDmult h_{square}$, where the objects are represented as the binding of attributes (blue, circle, yellow, square) and the scene a bundling of objects (blue circle, yellow square). By high-dimensional statistics, $h_{bcys}$ will share high similarity with the ``blue circle", $h_{blue}\HDmult h_{circle}$,  and ``yellow square" due to bundling and will share almost no similarity with any other colored shapes (or color and shape alone).

Using the basic HDC operators, one may construct composite hypervectors representing more complex data structures~\cite{rachkovskij2001binding,hersche2023neuro}, such as trees~\cite{frady2020resonator} and graphs~\cite{poduval2022graphd,nunes2022graphhd}. HDC's benefit lies in the interpretable hypervector manipulation: every transform by the HDC operator can be interpreted as a certain type of data structure manipulation~\cite{quiroz2020semantic,plate1995holographic}. As a result, decoding plays a crucial role in understanding and utilizing the information encoded within the hypervectors.

\subsection{Hypervector Factorization Problem} \label{subsec:Factorization}

One can decode bundled and permuted hypervectors based on their codebooks with reasonable efficiency and performance: bundling retains high similarity with constituents, making individual component retrieval straightforward, akin to set membership detection; Permutation involves a linear search algorithm, finding position and identity via permutations and codebook matches. 
Unlike bundling, binding lacks similarity retention, necessitating an exhaustive search over atomic hypervector combinations; and unlike permutation, the search space scales exponentially in the number of factors (resp. linearly w.r.t. sequence length), making the exhaustive search unfeasible. 
We formulate The Hypervector Factorization problem as follows~\cite{frady2020resonator}:
\begin{definition}
    Given $F$ codebooks $\mathcal{C}_1, \cdots, \mathcal{C}_F$ and a hypervector $c$, find $c_{1} \in \mathcal{C}_1, \cdots, c_F \in \mathcal{C}_F$ such that $c=\bigodot_{i=1}^{F} c_{i}.$
\end{definition}
In HDC, we assume that each codebook $\mathcal{C}_i$ is generated randomly and independently as in the MAP scheme. 
Consequently, the effective search space of the problem is $\Pi_{i = 1}^{F}|\mathcal{C}_i| = O(N^F)$. We denote the dimension of the hypervector as $D$.


\noindent \textbf{The capacity limitation of classical computing:} 
The problem of Hypervector Factorization can be formulated as an optimization problem and hence approached by optimization-based methods. In addition, \cite{frady2020resonator} leverages the principle of HDC and proposes the Resonator Network. When compared to optimization-based methods, it improves the algorithm's operational capacity at the cost of total convergence~\cite{kent2020resonator}.
Despite the added efficiency, classical methods still face the fundamental limit of the problem size, and this limit comes in two senses. First, the problem space scaled poorly w.r.t. the codebook size and the number of factors, and none of the classical approaches has shown sublinear scalability. 
Secondly, for the Resonator Network alone, the search space is also fundamentally limited by the dimension of the hyperspace. 

To overcome the problems of capacity and convergence and to improve upon efficiency, we introduce \Design, which leverages the power of quantum computing to perform hypervector factorization. It may provide a quadratic speedup as compared to classical search by massively alleviating the capacity issue. 

\section{Quantum Factorization} \label{sec:QFactor}
In this section, we introduce \Design which leverages the power of quantum computing to perform HD vector factorization efficiently. \Design uses a modified form of Grover's algorithm to search through all possible factorizations to find the correct one. Our main contribution is the modification to the state preparation operator to cater to the HDC use case and the design of the oracle that can identify the correct factorization. Grover's algorithm is a general technique for performing an unstructured search over a database and requires a state preparation algorithm and an oracle that identifies the correct solution. See the appendix for a short review of quantum computing basics. 

In \Design, we represent each element of our hypervector by a two-component qubit state: $\ket{0}$ will represent $1$ and $\ket{1}$ will represent $-1$. The multiplication of the bipolar components of the vector will be represented by the XOR of the binary labels of the corresponding states representing the hypervectors. We represent the $D-$dimensional hypervector $h_i$ by the corresponding qubit series $\ket{q^i_0q^i_1....q^i_{D-1}}\equiv \ket{\vec{q}^i}$. We can store multiple collections of hypervectors as $\ket{\vec{q}^1}\ket{\vec{q}^2}\cdot\cdot\cdot \ket{\vec{q}^n}$, which requires a total of $m\times D$ qubits to represent the $m$ hypervector factors. The binding of two HDC vectors $h_i \HDmult h_j$ is represented by $\ket{\vec{q}^i\oplus \vec{q}^j} \equiv \ket{(q^i_1\oplus q^j_1)(q^i_2\oplus q^j_2)....(q^i_{D-1}\oplus q^j_{D-1})}$. 

\subsection{Modified Grover's Algorithm}
Grover's search algorithm is based on the principles of amplitude amplification and phase inversion~\cite{brassard2002quantum,byrnes2018generalized}, leveraging the superposition and interference properties of quantum computing to perform an efficient search operation among a given list of items. The algorithm's fundamental objective is to find a specific item, often referred to as the target, within an unsorted database of size $S$. Grover's algorithm achieves this by iteratively applying a series of quantum operations to amplify the probability of measuring the target item. It achieves the quadratic speedup by reducing the search time from $O(S)$ to approximately $O(\sqrt{S})$ where $S$ is the total number of items.

The search operation consists of three steps: (1) The state preparation algorithm, (2) the Oracle and (3) the diffusion operator. The state preparation algorithm builds the operator $U_{\mathcal A}$ which generates a uniform superposition of all the possible states as
$U_{\mathcal{A}}\ket{\vec{0}} = \sum_{a\in \mathcal{A}} \frac{1}{\sqrt{N}}\ket{a}=\ket{\mathcal{A}},$ where $\mathcal A$ is the set of items we are searching over. The Oracle $U_o$ searches for the marked state (which we shall denote $\ket{m}$) with the action 
\begin{align}
    U_{o}\ket{a}=\begin{cases} -\ket{a} \hphantom{----}\ket{m}=\ket{a} \\ + \ket{a} \hphantom{----}\ket{m}\ne \ket{a}\end{cases}.
\end{align}
The diffusion operator $U_D$ inverts the amplitudes of the states in the codebook by the average, given by 
\begin{align}
    U_D=2\ket{\mathcal{A}}\bra{{\mathcal{A}}}-{I} = U_{\mathcal{A}}\left(2\ket{\vec{0}}\bra{\vec{0}}-I\right)U_{\mathcal{A}}^\dagger,
\end{align}
with $I$ the identity operator. Its action in a general superposition is given by 
\begin{align}
    U_D\sum_{a\in \mathcal{A}} \psi(a)\ket{a}=\sum_{a\in \mathcal{A}} \left(\mu-\psi(a)\right)\ket{a},
\end{align}
where $\psi(a)$ is the amplitude of state $\ket{a}$, and $\mu=\frac{1}{|\mathcal{A}|}\sum_{a\in\mathcal{A}}\psi(a)$ is the average of the amplitudes. 

Grover's algorithm can now be summarised as follows:
\begin{itemize}
    \item Construct the initial superposition of states given by $\ket{\mathcal{A}}$
    \item Apply the oracle operator to $U_o$, to get $U_o\ket{\mathcal{A}}$
    \item Apply the diffusion operator $U_D$, to get $U_DU_o\ket{\mathcal{A}}$
    \item Repeat the previous two steps $\frac{\pi}{4}\sqrt{S}$ number of times, and perform the measurements
\end{itemize}
After $\approx \frac{\pi}{4}\sqrt{S}$ iterations, the probability of measuring the marked state increases to a maximum value of about $50\%$, thus increasing the likelihood of finding the marked state.


In \Design, we represent each element of our hypervector by a two-component qubit state: $\ket{0}$ will represent $1$ and $\ket{1}$ will represent $-1$. The multiplication of the components of the vector will be represented by the XOR of the labels of the corresponding states representing the hypervectors. We represent the $D-$dimensional hypervector $h_i$ by the corresponding qubit series $\ket{q^i_0q^i_1....q^i_{D-1}}$, which we will refer to by the shorthand $ \ket{\vec{q}^i}$. We can store multiple collections of hypervectors as $\ket{\vec{q}^1}\ket{\vec{q}^2}\cdot\cdot\cdot \ket{\vec{q}^f}$, which requires a total of $f\times D$ qubits to represent the $f$ hypervector factors. The binding of two HDC vectors $h_i \HDmult h_j$ is represented by $\ket{\vec{q}^i\oplus \vec{q}^j} \equiv \ket{(q^i_1\oplus q^j_1)(q^i_2\oplus q^j_2)....(q^i_{D-1}\oplus q^j_{D-1})}$. The Oracle will need to implement the binding operation as part of its overall calculations. We will demonstrate a method to perform this that requires a number of gates that is linear to the dimension.

\begin{figure*}[h]
    \centering
    \includegraphics[width=0.9\textwidth]{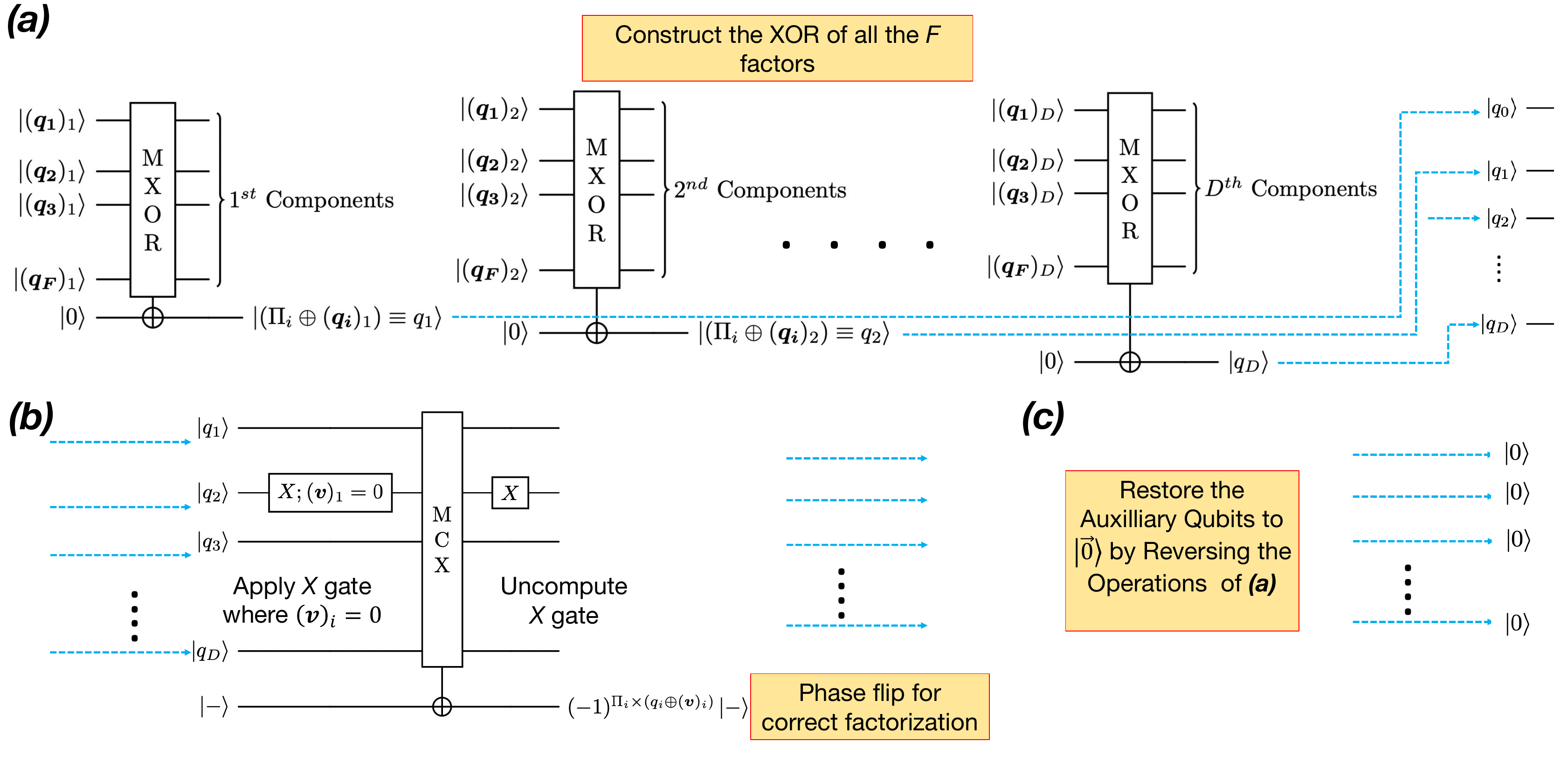}
    \caption{Construction of the Oracle circuit }
    \label{fig:oracle}
\end{figure*}

\subsection{State Preparation}
Given a specific location of the factor, the state preparation operator creates a superposition of all factors in the codebook. 
We assume that the factors belong to a codebook $M_{F\times N\times D}$, where $F$ is the number of factors, $N$ is the number of vectors that are candidate factors at each location, and $D$ is the dimension of the hypervector. Defining $\vec{q}^i_f$ to be the $i^{th}$ vector at the $f^{th}$ factor (i.e, $\left(\vec{q}^i_f\right)_d= M_{fid}$), the goal of the state preparation algorithm is to construct the following state: $U_{\mathcal{M}}\ket{\vec{0}}\cdot\cdot\cdot\ket{\vec{0}} = \frac{1}{\sqrt{N^F}}\left(\sum_i\ket{\vec{q}^i_1}\right)\cdot\cdot\cdot\left(\sum_i\ket{\vec{q}^i_F}\right)$, which represent all possible factorizations. 
This is implemented sequentially, by applying the algorithm of \cite{ventura1999initializing} to each factor individually by defining the operator $U_{\mathcal{M}}^f\ket{\vec{0}} = \frac{1}{\sqrt{N}}\left(\sum_i\ket{\vec{q}_f^i}\right)$,
which constructs a uniform superposition of all vectors belonging to the $k^{th}$ factor of the codebook. By defining the tensor product operator $U_{\mathcal{M}}= \Pi_f \otimes U_{\mathcal{M}}^f$, where each term acts on a different set of qubits as
\begin{align}
U_{\mathcal{M}}\ket{\vec{0}}\cdot\cdot\cdot\ket{\vec{0}} = \left(U^1_{\mathcal{M}}\ket{\vec{0}}\right)\left(U^2_{\mathcal{M}}\ket{\vec{0}}\right)\cdot\cdot\left(U^F_{\mathcal{M}}\ket{\vec{0}}\right),
\end{align}
we can generate a uniform superposition of all possible factors with a polynomial circuit complexity and qubit requirement. 

\subsection{Design of the Oracle}

We design our Oracle by implementing the operator $U_f$ which performs the operation $U_f\ket{\psi}\ket{c} = \ket{\psi}\ket{c\oplus 1}$, if $\ket{\psi}$ represents a correct factorization, and nothing otherwise. Here, $\ket{c}$ is a single-qubit output line. By initializing $c$ in the Hadamard state $\ket{c}=\ket{-}=\frac{{1}}{\sqrt{2}}(\ket{1}-\ket{0})$, $U_f$ implements the phase-kickback version of the oracle with the action defined previously. Given a binary vector $\vec{v}$, and a multi-qubit state $\ket{\vec{q}^1}\ket{\vec{q}^2}....\ket{\vec{q}^F}$ representing the factors, the oracle needs to check whether $\vec{q}^1\oplus \vec{q}^2\oplus ...\oplus \vec{q}^F=\vec{v}$, or alternatively if $\vec{q}^1\oplus \vec{q}^2\oplus ...\oplus \vec{q}^F\oplus \vec{v}=\vec{0}$. Our implementation of the Oracle circuit is illustrated in Fig.~\ref{fig:oracle}, which proceeds in three steps. 

First, as shown in Fig.~\ref{fig:oracle}(a), the circuit calculates the product of the $F$ qubit vectors $\vec{q}={\vec{q}^1\oplus...\oplus\vec{q}^F\oplus\vec{v}}$, performed component-wise using the ${\rm MXOR}$ operation, with the output loaded onto a set of auxilliary qubits. The $i^{th}$ component of $\vec{q}$ is calculated as 
\begin{align}
\left(...\right)\ket{\left(\vec{q}\right)_j} = {\rm MXOR}\left[\left(\ket{\left(\vec{q}^1\right)_j}\ket{\left(\vec{q}^2\right)_j}..\ket{\left(\vec{q}^F\right)_j}\right),\ket{0}\right]
\end{align}
with the last $\ket{0}$ being the output line of the ${\rm MXOR}$ operation. For example, given $\ket{\vec{q}_1}=\ket{10110}$ and $\ket{\vec{q}_2}=\ket{01101}$, the ${\rm MXOR}$ is calculated as  $\ket{\vec{q}}=\ket{10110\oplus 01101}= \ket{1001}$ 

Second, the Oracle circuit performs the check of whether $\vec{q}$ and $\vec{v}$ are equal, which is done using the fact that $\vec{q}\oplus\vec{\overline{v}}=\vec{1}$ if and only if the two vectors are equal, where the overline denotes the ${\rm NOT}$ operation, and $\vec{1}$ denotes the vector with all components $1$. Now, the XOR of $\vec{q}$ and $\vec{\overline{v}}$ is equivalent to simply inverting the component bit of $\vec{q}$ wherever the component of $\vec{v}$ is equal $0$. Thus, the Oracle simply applies an ${\rm X}$ gate to the components of $\ket{\vec{q}}$ where the corresponding component of $\ket{v}$ is $0$. Then, the Oracle checks whether the resulting vector is equal to $\vec{1}$, which uses the ${\rm MCX}$ operator controlled onto the output line initialized to the $\ket{-}$ Hadamard state. If $\vec{q}=\vec{v}$, then the phase of the state is inverted by $-1$, and nothing happens otherwise. Afterwards, the ${\rm X}$ gate is applied again to restore the $\ket{\vec{q}}$ state. For example, consider $\ket{\vec{q}}= \ket{1001}$, from the previous example. If the target state was $\ket{\vec{v}}= \ket{1101}$, then we flip all the bits in $\ket{\vec{q}}$ where the corresponding bits i $\ket{\vec{v}}$ are $0$. This gives us a new state $\ket{\vec{q}'}= \ket{1011}$. The ${\rm AND}$ of all the components are then $0$, sigifying that $\ket{\vec{q}}$ was not equal to the target state $\ket{\vec{v}}$. 

In the final step, the Oracle proceeds to restore the $\ket{\vec{q}}$ state to the $\ket{\vec{0}}$ state by the process of uncomputation, so that the ancillary qubits are restores to their original values. This is performed by simply repeating the first step, by performing the operation, $\left(...\right)\ket{0} = {\rm MXOR}\left[\left(\ket{\left(\vec{q}^1\right)_j}\ket{\left(\vec{q}^2\right)_j}..\ket{\left(\vec{q}^F\right)_j}\right),\ket{\left(\vec{q}\right)_j}\right],$ for each component $j=1,2,..,D$. This marks the construction of the Oracle. 


\begin{figure}[h]
    \centering
    \includegraphics[width=\columnwidth]{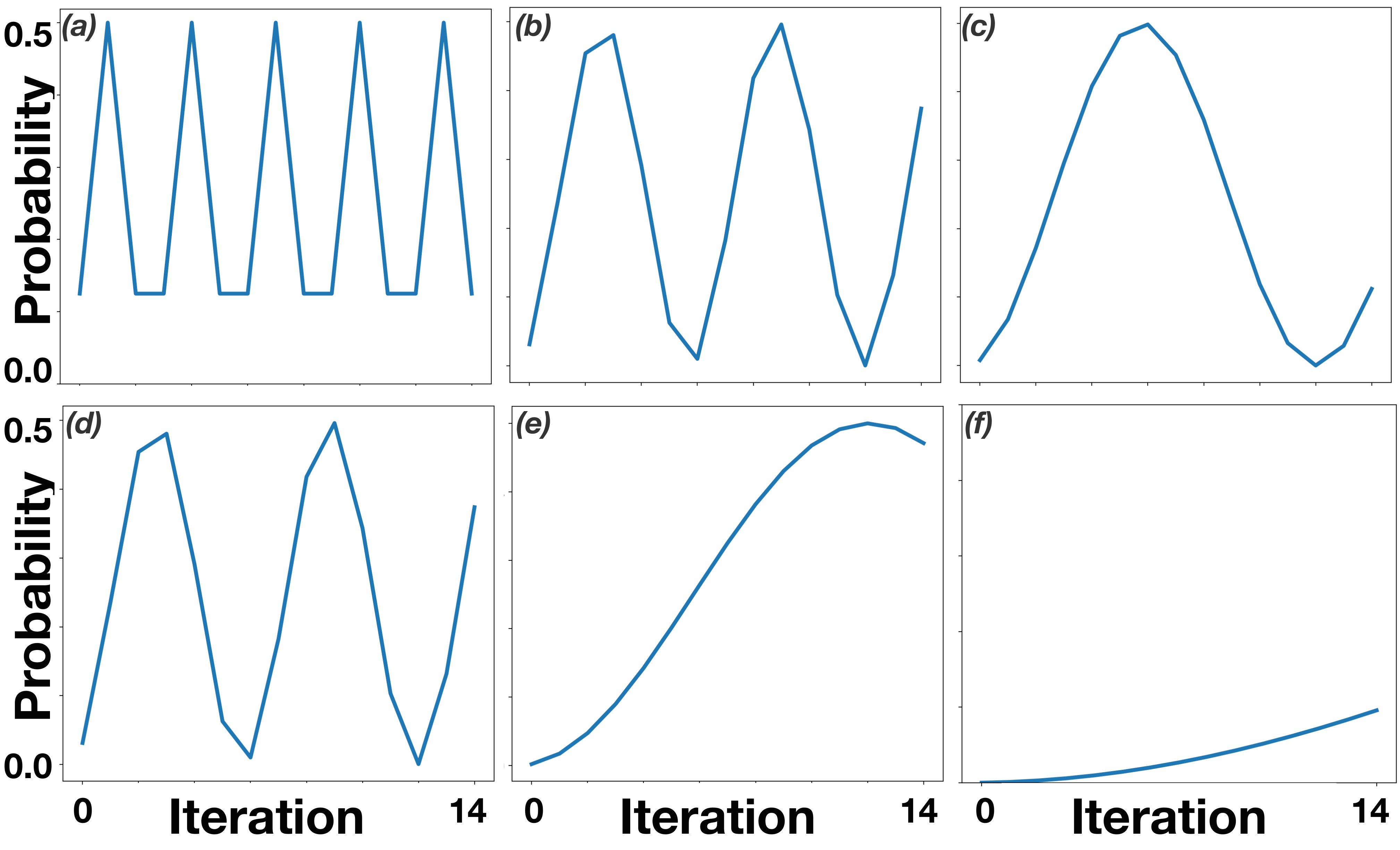}
    \caption{The probability of measuring the correct state as a function of the iteration for various parameters of the codebook size.  (a),(b) and (c), we perform factorization over $F=2$ factors, with $N=2,4$ and $8$ respectively. Similarly, in (d),(e) and (f), we factorize over $F=4$ factors with $N=2,4$ and $8$. The probability of measuring the correct factorization is an oscillating function of the iteration, which is an artifact of the amplitude inversion step.\vspace{3mm}}
    \label{fig:probabilityiter}
\end{figure}

\begin{figure}[h]
    \centering
    \includegraphics[width=\columnwidth]{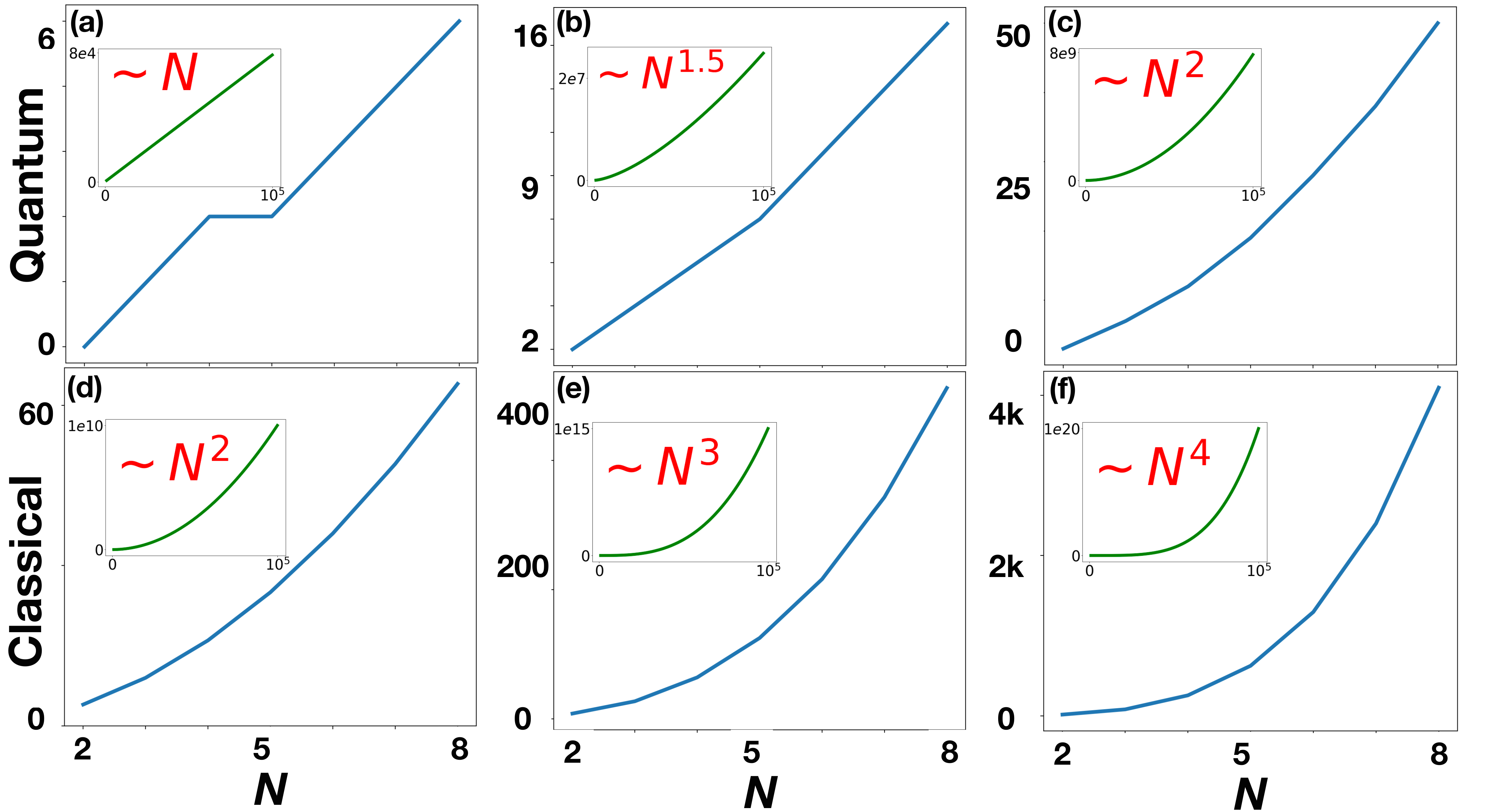}
    \caption{Scaling of the quantum and classical factorization complexity (a),(b) and (c) shows the scaling of the number of iterations required for the optimal measurement as a function of $N$ for $F=2,3$ and $4$ respectively. In the inset, we show projected scaling for larger values of $N$. (d)-(f), shows the classical complexity of a brute-force search for similar parameters as (a)-(c) respectively, illustrating the quadratic speedup gained from the quantum algorithm compared to the classical algorithm.\vspace{5mm}}
    \label{fig:quantumscaling}
\end{figure}

\section{Results} \label{sec:results}

\subsection{Factorization Performance} \label{subsec:performance}

In this section, we perform simulations of the modified Gover's circuit to understand the feasibility of factorization using a quantum computer. Using the state vector simulation method of qiskit \cite{Qiskit}, we implement the circuit and simulate its action to verify that the results correspond with the theory without any noise, by factorization over randomly generated codebooks (with $D=5$ dimensions). 

In Fig.~\ref{fig:probabilityiter}, we show the probability of measuring the correct state as a function of the iteration of each step in the Grover loop, for various parameters of the codebook size. In (a),(b) and (c), we factorize over $F=2$ factors, with $N=2,4$ and $8$ respectively. Similarly, in (d),(e) and (f), we factorize over $F=4$ factors with $N=2,4$ and $8$. We can see that the probability of measuring the correct factorization is an oscillating function of the iteration-- an artifact of the amplitude inversion step. In Fig.~\ref{fig:quantumscaling}(a)-(c), we show the scaling of the number of iterations required for the optimal measurement as a function of $N$ for $F=2,3$ and $4$ respectively. In the inset, we show projected scaling for larger values of $N$. For (d)-(f), we show the classical complexity of a brute-force search for similar parameters as (a)-(c) respectively. We can clearly observe the quadratic speedup gained from the quantum algorithm compared to the classical algorithm. The quantum complexity scales as $\sim N^{F/2}$, and the classical complexity goes as $\sim N^{F}$.



It is imperative to perform Grover's algorithm over the optimum number of iterations, which is given by $\frac{\pi}{4}\sqrt{N_{\rm tot}}$, where $N_{\rm tot}$ is the total number of possible states the algorithm is searching over. 
In our setup, $N_{\rm tot}=N^F$, assuming that the product of each factor in the codebook is unique. Therefore, the Grover's iteration needs to be stopped after $N^{F/2}$ iteration. From (a) to (c), as $nC$ increases the total number of combinations also increase, due to which the period of oscillations increase. A similar trend is seen from (d) to (f).

However, the preceeding analysis is contingent on the fact that the factorization is unique. If there are multiple different factorizations for the given hypervector within the codebook, then the peak amplitude will occur after $\left(N^F/t\right)^{1/2}$ iterations, where $t$ is the number of factorizations possible. The assumption that we usually make is that $t=1$ based on the randomness of hyperevectors for large dimensions. For small dimensions, the value of $t$ can be estimated using the solution counting algorithm, which is a variation of Grover's algorithm to estimate the number of solutions to a certain problem~\cite{brassard1998quantum}. 

\begin{figure}[h]
    \centering
    \includegraphics[width=\columnwidth]{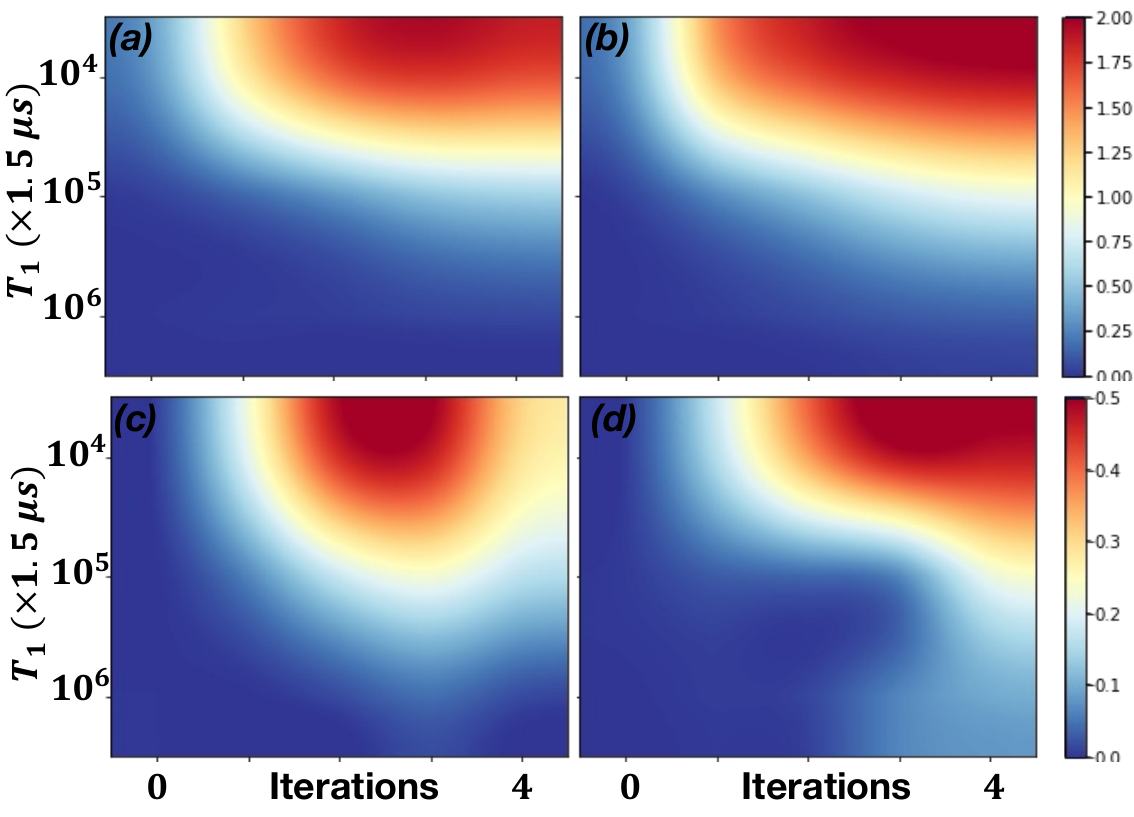}
    \caption{Error simulations in the presence thermal qubit relaxation times $T_1$ and $T_2$ with $T_2=2T_1$. (a) and (b) shows the error in the probability distributions as a whole due to qubit relaxation as a function of iterations in the Grover's algorithm and as a function of $T_1$ for $N=4$ and $N=5$ respectively. (c) and (d), shows the error in the probability of measuring the correct state for the same parameters as (a) and (b) respectively.}
    \label{fig:errorqc}
\end{figure}

\subsection{Factorizing with Noise} \label{subsec:noise}

Our current analysis is based on the assumption that there is no noise in the system. However, this is an ideal assumption far from reality since current quantum computers have a large amount of noise which manifest in different channels. However, we clarify that there are various types of overheads in a quantum device which needs to be resolved on the technical side in the future. These errors are primarly classified into (1) gate error on each qubit, (2) thermal relaxation error of qubits and (3) readout error for each qubit. (1) is primarily the error on the system during the application of each gate, and it depends on the type of implementation of the basis gates, and the number of qubits the gate acts on. (2) is simply the relaxation of qubits caused by thermal noise. This can result in an otherwise entangled state to lose coherence due to thermal relaxation of the qubit state, a process called decoherence. The process of decoherence is characterized by the timescale over which the qubit thermalizes, which are primarily the $T_1$ and $T_2$ relaxation timescale. 

In Fig.~\ref{fig:errorqc}, we show error simulations in the presence thermal qubit relaxation times $T_1$ and $T_2$ with $T_2=2T_1$. (a) and (b) shows the error in the probability distributions as a function of $T_1$ for $N=4$ and $N=5$ respectively. 
We observe that as $T_1$ reduces, the error in the state distribution increases in proportion since qubits lose coherence over a shorter timescale. The error also increases with the number of iterations since the system needs to be held in entanglement for a longer amount of time, which increases decoherence. (c) and (d) shows the error in the probability of measuring the correct state for the same parameters as (a) and (b) respectively. In (c), we note that as the iterations increases the error slightly decreases because of the periodic nature of Grover's algorithm. 

The error in quantum systems can be reduced in the future by developing systems with a longer relaxation timescale. Once scalability of qubits are achieved, then there are multiple error correction algorithms, in which the state of one \textit{logical} qubit is represented by an entangled state of multiple physical qubits, thus enabling added redundacy. In the near-term, there are multiple error \textit{mitigation} techniques, which try analyzing the patterns of error in the system to extrapolate a zero-error result. 

\begin{figure*}[h]
    \centering
     \includegraphics[width=\textwidth]{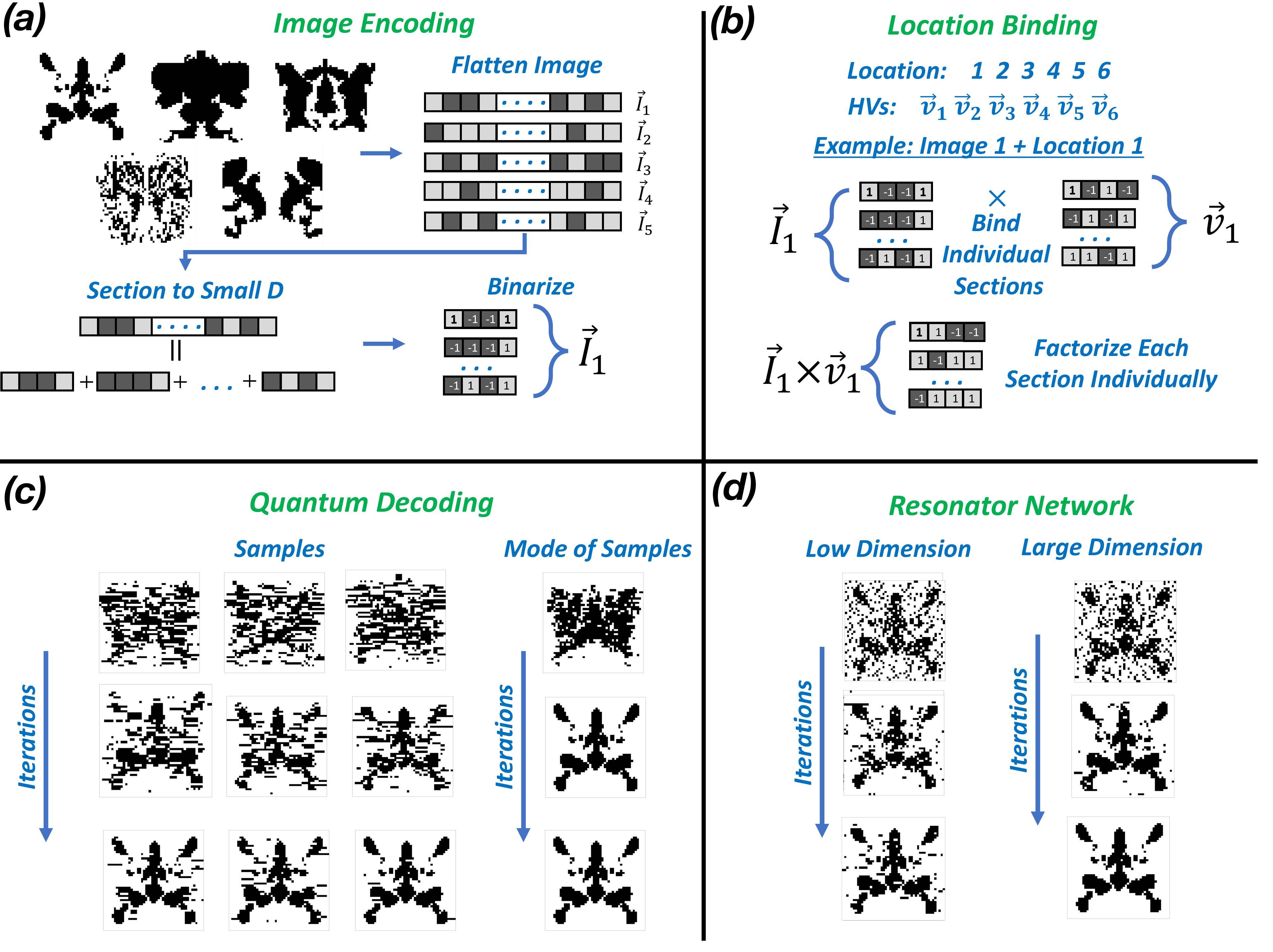}
     \caption{Image decoding. (a) Each image is polarized and sectioned into vectors of length $D$. (b) Location hypervectors are generated and sectioned similarly. An image and its location are associated via binding. (c) Performance for quantum decoding using \Design. (d) Performance for the resonator network in low dimension and high dimension settings.} 
   \label{fig:case1}
 \end{figure*}

\subsection{Case Study: Factorization Capacity and Image-Location Decoding} \label{subsec:case1}

In this case study, we compare the performance of \Design to that of the Resonator Network and demonstrate the difference in capacity limitation between the quantum and classical settings.

Figure \ref{fig:case1} demonstrates the factorization accuracy for image sequence decoding, where we associate each image to a location in the sequence. 
Given a set of $48 \times 48$ binary images, we polarize each image (such that the $0$-entries becomes $-1$), and flatten it to a vector $\vec{I}$. The list of vectors constitutes the image codebook $\mathcal{I}$ [\ref{fig:case1}(a)]. To capture sequencing with the binding operation, we generate random vectors $\vec{v}_1, ...,\vec{v}_L$ of dimension $48^2 = 2304$ as location codebook $\mathcal{L}$, where $L$ is the number of locations. 
To memorize the image $i$ at location $j$, we perform the binding operation $\vec{I}_i \HDmult \vec{v}_j$ [\ref{fig:case1}(b)]. In our simulations, we perform decoding by sectioning it evenly to small vectors of dimension $D=6$. 
We may then apply \Design to each section for decoding the problem. 

Figure \ref{fig:case1}(c) visualizes the recovered images for different iterations of \Design applied. The results align with our expectation from Figure \ref{fig:probabilityiter}. The probability of correctly recovering the image vectors increases with each iteration until the highest probability is reached, and since the probability of measuring the correct state is rarely $100\%$, some sections of the image are not recovered correctly but can be corrected by taking the statistical mode over multiple runs of the simulation. Therefore, \Design correctly recovers the images. 

In Figure \ref{fig:case1}(d) we show the resonator network with two settings. In the low dimension setting, the hypervector dimension is set to be the same as \Design; we also experiment with the large dimension setting, where a higher dimensional vector is used to encode each section of vectors. The results show that, in low dimension setting, the resonator network cannot always converge to the correct sections of the images. This is largely due to the signal-to-noise ratio demand mentioned in Section \ref{subsec:Factorization}. When this demand is met, as in the large dimension setting, the resonator network accurately retrieves the images.

\begin{figure*}[h]
   \centering
    \includegraphics[width=0.8\textwidth]{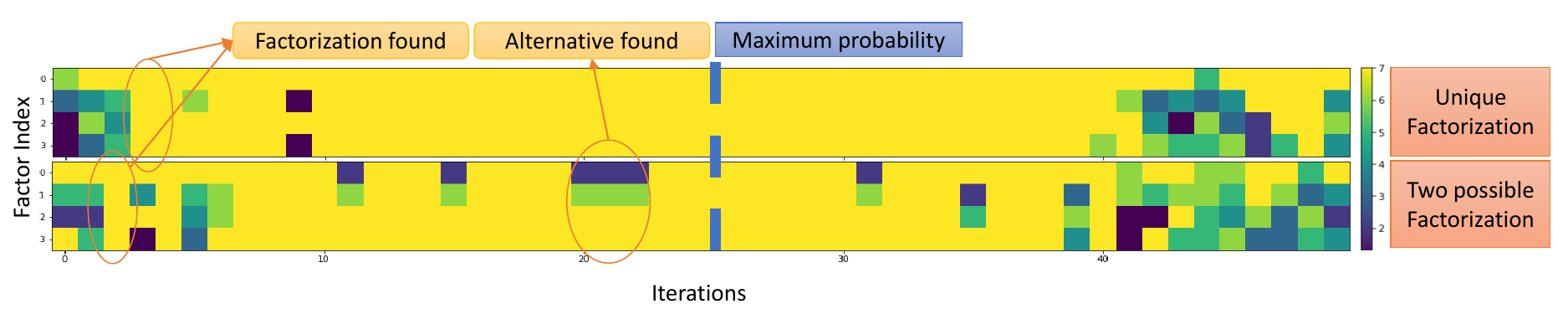}
    \caption{\Design behavior for non-unique factorization problem.} 
  \label{fig:case2}
\end{figure*}
\subsection{Case Study: Non-unique factorization} \label{subsec:case2}

\begin{table}[]
\setlength\tabcolsep{3pt}
\centering
\begin{tabular}{||l||llll||l||}\hline
$(D,F,N)$ & \multicolumn{4}{c||}{Resonator Network~\cite{frady2020resonator}}        & \multicolumn{1}{c||}{Quantum} \\\hline
         & $P_s$ & $N_{I}$ & $P_f$ & $N_{S}$ &  $N_{S}$ \\
$(100,3,10) $  & 0.33    & 17.21  & 0.9         & 52.15 & \textbf{24}   \\
$(100,4,10) $  & 0.0    & 9.0 & 0.96        & $ O(nC^N)$ & \textbf{78}    \\
$(25,3,5)$ & 0.12    & 7.83  & 0.16         & 65.25      &  \textbf{8}          \\
$(25,4,5)$ & 0.02    & 769.5  & 0.52        & 3475 & \textbf{19} \\\hline
\end{tabular}
\caption{Comparison of quantum search with resonator network, highlighting the optimality of quantum search. $P_s$ is the probability the resonator network converges to correct factorization, and $N_{I}$ is the average number of iterations required. $N_{S}=N_{I}/P_s$ is the effective number of steps required to find the correct result. $P_f$ is the probability of non-convergence for the resonator network. Note that $P_f$ is not the complement of $P_s$, because there are situations where the resonator network converges onto an incorrect factorization which is considered in neither $P_s$ or $P_f$. For quantum search, the number of steps required is effectively the number of iterations, since the maximum probability of finding the marked state is $\sim 1$.}
\label{tab:comparison}
\end{table}

The result from the previous section implies that since the problem size is no longer limited by the signal-to-noise ratio, the complexity of the problem now relies only on the problem size. Thus, we now have a huge improvement in the size of the allowed codebook for a fixed dimension using the corresponding quantum algorithm. In this section, we further study the behavior of \Design when this advantage is pushed to the limit: when the size of the combinations of factors exceeds that of the vector space, resulting in more than one solution. 

Figure \ref{fig:case2} visualizes the behavior of \Design for factorization problems when there is unique factorization and when there are two possible solutions. In both cases, we have $F=4$ number of factors and codebooks of size $N=7$ with hyperdimension $D=10$. Since $7^4 = 2401 > 2^{10} = 1024$, the pigeonhole principle guarantees that there are bound hypervectors with non-unique factorization. We find hypervectors with $1$ and $2$ factorization as the target hypervector respectively. 

As shown in Figure \ref{fig:case2}, \Design is able to recover (one of) the correct solution(s) in both cases at maximum probability. In fact, the correct is maintained within a relatively large region; the algorithm reaches the correct factorization at $4^{th}$ and $3^{rd}$ iteration respectively, and is relatively stable until around $40^{th}$ iteration. The region of iterations over which the correct factorization is recovered depends on the number of trials.
Indeed, further experiment shows that when we lower the number of trials for each case, the range of the ``correct region" reduces on both ends, and more fluctuation is observed close to the boundary. Though for practical concerns, the main goal is to run \Design to the optimum number of iterations, so that the number of measurements required is minimal. 

Interestingly, when there are $2$ solutions for the target hypervector, \Design fluctuates between the two. As shown in the bottom plot of Figure \ref{fig:case2}, after \Design finds the first solution, it changes to the alternative solution for certain iterations. This cannot be explained away by quantum measurement, as fluctuation appears and is maintained for several iterations close to the $25^{th}$ one. 

In Table.~\ref{tab:comparison}, we show the effective umber of steps required for decoding by both the resonator network and quatum method. The quatum method is show to be superior, since it is not effected by the presence of cross correlatio noise, and always decodes correctly, eventually.

\section{Conclusion}

This introduces a quantum computing algorithm designed to address the Hypervector Factorization Problem, a key decoding problem in interpretable HDC models for learning and cognitive processing. Our qualitative results show that \Design provides a quadratic speedup over the search and is not restricted to the capacity limit suffered by classical algorithms. Looking forward, we anticipate that our research will inspire further investigations into quantum algorithms optimized for Hyperdimensional Computing. We envisage that future work will involve hybrid approaches, integrating classical and quantum computing methods to harness the unique strengths of both worlds, further enhancing the efficiency and performance of HDC systems.

\textbf{Limitations: } The main limitation of our work is the applicability on actual quantum hardware. Today's technology suffers from lack of available coherent qubits and lack of error correction, while our method works in high dimensional vector spaces, requiring a large number of qubits. However, we foresee that this might be available in the near future. A new local error correction code can create $100$ logical qubits using only $\sim 1000$ physical qubits with an error of $\sim 10^{-8}$~\cite{ruiz2024ldpc}. Moreover, HDC does not require such accuracy (since we rely on approximate orthogonality), and as such we could use around $5000$ physical qubits to represent $1500$ HDC qubits, which is enough to perform $3-$ factorization with $D=500$. With new technological advancements in quantum computing, our research direction opens a new avenue for neurosymbolic quantum computation.



\section*{Acknowledgments}
This work was supported in part by the DARPA Young Faculty Award, the National Science Foundation (NSF) under Grants \#2127780, \#2319198, \#2321840, \#2312517, and \#2235472, the Semiconductor Research Corporation (SRC), the Office of Naval Research through the Young Investigator Program Award, and Grants \#N00014-21-1-2225 and N00014-22-1-2067. Additionally, support was provided by the Air Force Office of Scientific Research under Award \#FA9550-22-1-0253, along with generous gifts from Xilinx and Cisco.



\bibliography{mybibliography,CLOG}

\begin{thebibliography}{30}
\providecommand{\natexlab}[1]{#1}
\providecommand{\url}[1]{\texttt{#1}}
\expandafter\ifx\csname urlstyle\endcsname\relax
  \providecommand{\doi}[1]{doi: #1}\else
  \providecommand{\doi}{doi: \begingroup \urlstyle{rm}\Url}\fi

\bibitem[Brassard et~al.(1998)Brassard, H{\o}yer, and
  Tapp]{brassard1998quantum}
G.~Brassard, P.~H{\o}yer, and A.~Tapp.
\newblock Quantum counting.
\newblock In \emph{Automata, Languages and Programming: 25th International
  Colloquium, ICALP'98 Aalborg, Denmark, July 13--17, 1998 Proceedings 25},
  pages 820--831. Springer, 1998.

\bibitem[Brassard et~al.(2002)Brassard, Hoyer, Mosca, and
  Tapp]{brassard2002quantum}
G.~Brassard, P.~Hoyer, M.~Mosca, and A.~Tapp.
\newblock Quantum amplitude amplification and estimation.
\newblock \emph{Contemporary Mathematics}, 305:\penalty0 53--74, 2002.

\bibitem[Byrnes et~al.(2018)Byrnes, Forster, and
  Tessler]{byrnes2018generalized}
T.~Byrnes, G.~Forster, and L.~Tessler.
\newblock Generalized grover’s algorithm for multiple phase inversion states.
\newblock \emph{Physical review letters}, 120\penalty0 (6):\penalty0 060501,
  2018.

\bibitem[Frady et~al.(2018)Frady, Kleyko, and Sommer]{frady2018theory}
E.~P. Frady, D.~Kleyko, and F.~T. Sommer.
\newblock A theory of sequence indexing and working memory in recurrent neural
  networks.
\newblock \emph{Neural Computation}, 30\penalty0 (6):\penalty0 1449--1513,
  2018.

\bibitem[Frady et~al.(2020)Frady, Kent, Olshausen, and
  Sommer]{frady2020resonator}
E.~P. Frady, S.~J. Kent, B.~A. Olshausen, and F.~T. Sommer.
\newblock Resonator networks, 1: an efficient solution for factoring
  high-dimensional, distributed representations of data structures.
\newblock \emph{Neural computation}, 32\penalty0 (12):\penalty0 2311--2331,
  2020.

\bibitem[Gayler(1998)]{gayler1998multiplicative}
R.~W. Gayler.
\newblock Multiplicative binding, representation operators \& analogy (workshop
  poster).
\newblock 1998.

\bibitem[Hersche et~al.(2023)Hersche, Zeqiri, Benini, Sebastian, and
  Rahimi]{hersche2023neuro}
M.~Hersche, M.~Zeqiri, L.~Benini, A.~Sebastian, and A.~Rahimi.
\newblock A neuro-vector-symbolic architecture for solving raven’s
  progressive matrices.
\newblock \emph{Nature Machine Intelligence}, 5\penalty0 (4):\penalty0
  363--375, 2023.

\bibitem[Imani et~al.(2022)Imani, Zakeri, Chen, Kim, Poduval, Lee, Kim,
  Sadredini, and Imani]{imani2022neural}
M.~Imani, A.~Zakeri, H.~Chen, T.~Kim, P.~Poduval, H.~Lee, Y.~Kim, E.~Sadredini,
  and F.~Imani.
\newblock Neural computation for robust and holographic face detection.
\newblock In \emph{Proceedings of the 59th ACM/IEEE Design Automation
  Conference}, pages 31--36, 2022.

\bibitem[Kanerva(2009)]{kanerva2009hyperdimensional}
P.~Kanerva.
\newblock Hyperdimensional computing: An introduction to computing in
  distributed representation with high-dimensional vectors.
\newblock \emph{Cognitive Computation}, 2009.

\bibitem[Kent et~al.(2020)Kent, Frady, Sommer, and
  Olshausen]{kent2020resonator}
S.~J. Kent, E.~P. Frady, F.~T. Sommer, and B.~A. Olshausen.
\newblock Resonator networks, 2: Factorization performance and capacity
  compared to optimization-based methods.
\newblock \emph{Neural computation}, 32\penalty0 (12):\penalty0 2332--2388,
  2020.

\bibitem[Kim et~al.(2018)Kim, Imani, and Rosing]{kim2018efficient}
Y.~Kim, M.~Imani, and T.~S. Rosing.
\newblock Efficient human activity recognition using hyperdimensional
  computing.
\newblock In \emph{Proceedings of the 8th International Conference on the
  Internet of Things}, pages 1--6, 2018.

\bibitem[Kleyko et~al.(2023)Kleyko, Rachkovskij, Osipov, and
  Rahimi]{kleyko2023survey}
D.~Kleyko, D.~Rachkovskij, E.~Osipov, and A.~Rahimi.
\newblock A survey on hyperdimensional computing aka vector symbolic
  architectures, part ii: Applications, cognitive models, and challenges.
\newblock \emph{ACM Computing Surveys}, 55\penalty0 (9):\penalty0 1--52, 2023.

\bibitem[Ni et~al.(2023)Ni, Chen, Poduval, Zou, Mercati, and
  Imani]{ni2023brain}
Y.~Ni, H.~Chen, P.~Poduval, Z.~Zou, P.~Mercati, and M.~Imani.
\newblock Brain-inspired trustworthy hyperdimensional computing with efficient
  uncertainty quantification.
\newblock In \emph{2023 IEEE/ACM International Conference on Computer Aided
  Design (ICCAD)}, pages 01--09. IEEE, 2023.

\bibitem[Nunes et~al.(2022)Nunes, Heddes, Givargis, Nicolau, and
  Veidenbaum]{nunes2022graphhd}
I.~Nunes, M.~Heddes, T.~Givargis, A.~Nicolau, and A.~Veidenbaum.
\newblock Graphhd: Efficient graph classification using hyperdimensional
  computing.
\newblock In \emph{2022 Design, Automation \& Test in Europe Conference \&
  Exhibition (DATE)}, pages 1485--1490. IEEE, 2022.

\bibitem[Plate(1995)]{plate1995holographic}
T.~A. Plate.
\newblock Holographic reduced representations.
\newblock \emph{IEEE Transactions on Neural networks}, 6\penalty0 (3):\penalty0
  623--641, 1995.

\bibitem[Poduval et~al.(2021{\natexlab{a}})Poduval, Issa, Imani, Zhuo, Yin,
  Najafi, and Imani]{poduval2021robust}
P.~Poduval, M.~Issa, F.~Imani, C.~Zhuo, X.~Yin, H.~Najafi, and M.~Imani.
\newblock Robust in-memory computing with hyperdimensional stochastic
  representation.
\newblock In \emph{2021 IEEE/ACM International Symposium on Nanoscale
  Architectures (NANOARCH)}, pages 1--6. IEEE, 2021{\natexlab{a}}.

\bibitem[Poduval et~al.(2021{\natexlab{b}})Poduval, Ni, Kim, Ni, Kumar,
  Cammarota, and Imani]{poduval2021hyperdimensional}
P.~Poduval, Y.~Ni, Y.~Kim, K.~Ni, R.~Kumar, R.~Cammarota, and M.~Imani.
\newblock Hyperdimensional self-learning systems robust to technology noise and
  bit-flip attacks.
\newblock In \emph{IEEE/ACM International Conference on Computer-Aided Design
  (ICCAD)}, 2021{\natexlab{b}}.

\bibitem[Poduval et~al.(2021{\natexlab{c}})Poduval, Zou, Najafi, Homayoun, and
  Imani]{poduval2021stochd}
P.~Poduval, Z.~Zou, H.~Najafi, H.~Homayoun, and M.~Imani.
\newblock Stochd: Stochastic hyperdimensional system for efficient and robust
  learning from raw data.
\newblock In \emph{IEEE/ACM Design Automation Conference (DAC)},
  2021{\natexlab{c}}.

\bibitem[Poduval et~al.(2021{\natexlab{d}})Poduval, Zou, Yin, Sadredini, and
  Imani]{poduval2021cognitive}
P.~Poduval, Z.~Zou, X.~Yin, E.~Sadredini, and M.~Imani.
\newblock Cognitive correlative encoding for genome sequence matching in
  hyperdimensional system.
\newblock In \emph{IEEE/ACM Design Automation Conference (DAC)},
  2021{\natexlab{d}}.

\bibitem[Poduval et~al.(2022{\natexlab{a}})Poduval, Ni, Kim, Ni, Kumar,
  Cammarota, and Imani]{poduval2022adaptive}
P.~Poduval, Y.~Ni, Y.~Kim, K.~Ni, R.~Kumar, R.~Cammarota, and M.~Imani.
\newblock Adaptive neural recovery for highly robust brain-like representation.
\newblock In \emph{Proceedings of the 59th ACM/IEEE Design Automation
  Conference}, pages 367--372, 2022{\natexlab{a}}.

\bibitem[Poduval et~al.(2022{\natexlab{b}})Poduval, Zakeri, Imani, Alimohamadi,
  and Imani]{poduval2022graphd}
P.~Poduval, A.~Zakeri, F.~Imani, H.~Alimohamadi, and M.~Imani.
\newblock Graphd: Graph-based hyperdimensional memorization for brain-like
  cognitive learning.
\newblock \emph{Frontiers in Neuroscience}, page~5, 2022{\natexlab{b}}.

\bibitem[Poduval et~al.(2024)Poduval, Ni, Zou, Ni, and Imani]{poduval2024nethd}
P.~P. Poduval, Y.~Ni, Z.~Zou, K.~Ni, and M.~Imani.
\newblock Nethd: Neurally inspired integration of communication and learning in
  hyperspace.
\newblock \emph{Advanced Intelligent Systems}, page 2300841, 2024.

\bibitem[{Qiskit contributors}(2023)]{Qiskit}
{Qiskit contributors}.
\newblock Qiskit: An open-source framework for quantum computing, 2023.

\bibitem[Quiroz-Mercado et~al.(2020)Quiroz-Mercado, Barr{\'o}n-Fern{\'a}ndez,
  and Ram{\'\i}rez-Salinas]{quiroz2020semantic}
J.~I. Quiroz-Mercado, R.~Barr{\'o}n-Fern{\'a}ndez, and M.~A.
  Ram{\'\i}rez-Salinas.
\newblock Semantic similarity estimation using vector symbolic architectures.
\newblock \emph{IEEE Access}, 8:\penalty0 109120--109132, 2020.

\bibitem[Rachkovskij and Kussul(2001)]{rachkovskij2001binding}
D.~A. Rachkovskij and E.~M. Kussul.
\newblock Binding and normalization of binary sparse distributed
  representations by context-dependent thinning.
\newblock \emph{Neural Computation}, 13\penalty0 (2):\penalty0 411--452, 2001.

\bibitem[Rahimi et~al.(2017)Rahimi, Tchouprina, Kanerva, Mill{\'a}n, and
  Rabaey]{rahimi2017hyperdimensional2}
A.~Rahimi, A.~Tchouprina, P.~Kanerva, J.~d.~R. Mill{\'a}n, and J.~M. Rabaey.
\newblock Hyperdimensional computing for blind and one-shot classification of
  eeg error-related potentials.
\newblock \emph{Mobile Networks and Applications}, pages 1--12, 2017.

\bibitem[Ruiz et~al.(2024)Ruiz, Guillaud, Leverrier, Mirrahimi, and
  Vuillot]{ruiz2024ldpc}
D.~Ruiz, J.~Guillaud, A.~Leverrier, M.~Mirrahimi, and C.~Vuillot.
\newblock Ldpc-cat codes for low-overhead quantum computing in 2d.
\newblock \emph{arXiv preprint arXiv:2401.09541}, 2024.

\bibitem[Ventura and Martinez(1999)]{ventura1999initializing}
D.~Ventura and T.~Martinez.
\newblock Initializing the amplitude distribution of a quantum state.
\newblock \emph{Foundations of Physics Letters}, 12:\penalty0 547--559, 1999.

\bibitem[Yeung et~al.(2024)Yeung, Poduval, and Imani]{yeung2024self}
C.~Yeung, P.~Poduval, and M.~Imani.
\newblock Self-attention based semantic decomposition in vector symbolic
  architectures.
\newblock \emph{arXiv preprint arXiv:2403.13218}, 2024.

\bibitem[Zou et~al.(2022)Zou, Chen, Poduval, Kim, Imani, Sadredini, Cammarota,
  and Imani]{zou2022biohd}
Z.~Zou, H.~Chen, P.~Poduval, Y.~Kim, M.~Imani, E.~Sadredini, R.~Cammarota, and
  M.~Imani.
\newblock Biohd: an efficient genome sequence search platform using
  hyperdimensional memorization.
\newblock In \emph{Proceedings of the 49th Annual International Symposium on
  Computer Architecture}, pages 656--669, 2022.

\end{thebibliography}

\end{document}